\documentclass[preprint,aps,pra,showpacs,floatfix]{revtex4}
\usepackage{graphicx}
\usepackage{times}
\usepackage{nicefrac}
\usepackage{amsmath}
\usepackage{amsfonts}
\usepackage{amssymb}
\usepackage{amsthm}
\usepackage{epsf}
\usepackage{bm}
\usepackage{bbm}
\usepackage{times}
\usepackage[english]{babel}

\newcommand{\dE}{\Delta E}

\newcommand{\az}{\alpha Z}
\newcommand{\be}{\beta_{20}}
\newcommand{\bee}{\beta_{40}}
\newcommand{\albi}{\boldsymbol{\alpha}_i}

\newcommand{\rb}{\vec{r}}

\newcommand{\la}{\langle}
\newcommand{\ra}{\rangle}

\begin{document}
\title{Nuclear deformation effect on the binding energies in heavy ions
}
\date{}
\author{ Y.~S. ~Kozhedub$^1$, O.~V. ~Andreev$^1$,  V.~M. ~Shabaev$^{1,2}$, I.~I. Tupitsyn$^1$, C. Brandau$^3$, C. Kozhuharov$^3$, G. Plunien$^4$, and T. St\"{o}hlker$^{3,5}$}

\affiliation{
$^1$
Department of Physics, St.Petersburg State University, Oulianovskaya 1, Petrodvorets, St.Petersburg 198504, Russia \\
$^2$
Max-Planck-Institut f\"ur Physik Komplexer Systeme, N\"othnitzer Str. 38, D-01187 Dresden, Germany\\
$^3$
Gesellschaft f\"{u}r Schwerionenforschung (GSI), Planckstrasse 1, D-64291 Darmstadt, Germany\\
$^4$
Institut f\"{u}r Theoretische Physik, TU Dresden, Mommsenstrasse 13, D-01062 Dresden, Germany\\
$^5$
Physikalisches Institut, Ruprecht-Karls-Universitat Heidelberg, Philosophenweg 12, D-69120 Heidelberg, Germany\\
}
\begin{abstract}
Nuclear deformation effects on the binding energies in heavy ions are
investigated.
Approximate formulas for the nuclear-size correction and the isotope shift  for deformed nuclei
are derived. Combined with direct numerical evaluations, these formulas are employed to reanalyse experimental data on the nuclear-charge-distribution parameters in $^{238}\textrm{U}$ and to revise the nuclear-size corrections to the binding energies in H- and Li-like $^{238}\textrm{U}$. As a result, the theoretical uncertainties for the ground-state Lamb shift in $^{238}\textrm{U}^{91+}$ and for the $2p_{1/2}-2s$ transition energy in $^{238}\textrm{U}^{89+}$ are significantly reduced. The isotope shift of the $2p_{j}-2s$ transition energies for $^{142}\textrm{Nd}^{57+}$ and $^{150}\textrm{Nd}^{57+}$ is also evaluated including nuclear size and nuclear recoil effects within a full QED treatment.
\end{abstract}


\maketitle

%
%
%
\section{Introduction}

As is known (see, e.g., Refs \cite{Jo85,Sh93}), the finite-nuclear-size correction to the atomic energy
levels is sufficiently well determined by
the root-mean-square (rms) radius of the nucleus.
Following  Franosch and Soff \cite{Fr91}, the uncertainty due to this correction was usually estimated by adding quadratically two errors, one obtained by varying the rms radius within its error bar and the other obtained by changing the model of the nuclear charge distribution from the Fermi to the homogeneously-charged-sphere model. This rather conservative estimate was sufficient in so far as
the total theoretical uncertainty was mainly determined by other contributions.
The recent progress made
in calculations of higher-order QED and electron-correlation
corrections \cite{Yer03,Art05,Sha06,Ye06,Sh06a} and
 the current status of precision experiments with heavy few-electron ions
\cite{Sc91,El96,El98,Br03,Gum05,Be05,Brpu} require, however,
 a more accurate consideration of the nuclear shape and deformation effects.
Corresponding improvements are given in the present paper.

The finite-nuclear-size correction is studied both numerically and analytically. Approximate analytical formulas for this effect are derived in the general case of a deformed nucleus.
Special attention is paid to evaluation of the nuclear-size correction
to the binding energies of H- and Li-like uranium, where the most accurate experimental data were recently reported \cite{Gum05,Be05}. The study performed in the paper is employed to revise the value of the nuclear rms charge radius for $^{238}\textrm{U}$ and to recalculate the corresponding correction to the binding energies. As a result, the theoretical accuracy of the ground-state Lamb shift in $^{238}\textrm{U}^{91+}$ and of the $2p_{1/2}-2s$ transition energy in $^{238}\textrm{U}^{89+}$ are significantly improved.

The isotope shift of the $2p_{j}-2s$ transition energies for the isotopes A=142 and A=150 of Li-like $^\textrm{A}\textrm{Nd}^{57+}$ is evaluated as a function of the difference $\delta \la r^2 \ra$ of
 the nuclear mean-square  charge radius. The calculation includes the nuclear size correction to the one-electron Dirac binding energy as well as the corresponding effect on the electron-correlation, Breit interaction, and QED contributions. The mass shift including the nonrelativistic, relativistic, and QED recoil effects is also evaluated. Combined with an estimate of the nuclear polarization effect on the binding energy, these data can be used to extract the $\delta \la r^2 \ra$ value from the corresponding experiment \cite{Brpu}.

The relativistic unit system ($\hbar=c=m=1~)$ and the Heaviside charge unit $(\alpha=e^2/4\pi, e<0)$ are employed throughout the paper.

\section{Formulation}

The Coulomb interaction between an atomic electron and the nucleus is given by
\begin{align}\label{v1}
V(\rb_e, \rb_1, \dots , \rb_Z)
=-\frac{e^2}{4\pi}\sum^{Z}_{i=1}\frac{1}{|\rb_e-\rb_i|},
\end{align}
where $\rb_e$ is the electron position, $\rb_i$ is the
position of i-th proton, and the summation runs over all protons of the nucleus. If we neglect nuclear polarization effects, we can restrict our consideration of the operator $V$ to a model space, where the nuclear states may differ from each other only by the projection of the total angular momentum on the laboratory $Z$ axis.

In what follows, we assume that the nuclear Hamiltonian can be separated into rotational and intrinsic parts, the nucleus is axially symmetric and has reflection symmetry with respect to the plane which is perpendicular to the axial-symmetry axis. With this assumption, the nuclear wave function can be written as  \cite{Si72,ring_80}
\begin{align}\label{nwv1}
\vert IMK \ra=\sqrt{\frac{2I+1}{16\pi^2}} \bigg( D_{KM}^I
(\Phi,\Theta,\Psi) \chi_K^\lambda (\tau') + (-1)^{I-J}D_{-KM}^I
(\Phi,\Theta,\Psi) \chi_{-K}^\lambda (\tau') \bigg)
\end{align}
for $K\ne 0$ and
\begin{align}\label{nwv2}
\vert IM0 \ra=\sqrt{\frac{2I+1}{8\pi^2}} D_{0M}^I (\Phi,\Theta,\Psi)
\chi_0^\lambda (\tau')=\frac{1}{\sqrt{2\pi}}Y_{IM}(\Theta,\Phi)\chi_0^\lambda
(\tau')
\end{align}
for $K=0$, where $I$ is the total nuclear angular momentum, $M$ and $K$ are its projections on the laboratory and the nuclear body-fixed $Z$ axis, respectively, $\lambda$ denotes the other intrinsic quantum numbers,
and $(-1)^J$ must be considered as an operator defined by its action on the wave functions for given
intrinsic angular momenta \cite{Si72,ring_80}.
Here and below the prime indicates variables taken in the nuclear coordinate frame and $\tau'$ denotes the whole set of the internal nuclear coordinates.
The Euler angles $\Phi,\Theta,\Psi$ in the Wigner $D$-functions give the orientation of the intrinsic body-fixed
system with respect to the laboratory frame.

For a given internal nuclear state with $K=0$, we should average the
interaction operator $V(\rb_e, \rb_1, \dots , \rb_Z)$ with the internal nuclear
wave function $\chi_0^\lambda (\tau')$. We obtain
\begin{align}\label{v4}
\la \chi_0^{\lambda}| V | \chi_0^{\lambda}\ra =-\frac{e^2 Z}{4\pi}\int d\rb'
\frac{\rho(\rb')}
 {|\rb_e-\rb|}\,,
\end{align}
where the nuclear charge distribution  $\rho(\rb')$ is defined by
\begin{align}\label{pho0}
Z \rho(\vec r')=\la \chi_0^{\lambda}| \sum_{i=1}^{Z} \delta(\rb'-\rb_i')| \chi_0^{\lambda}\ra\,
\end{align}
and $\rb$ denotes the position of the $\rb'$ vector in the laboratory coordinate frame.
With the assumptions considered above, the density $\rho(\vec{r}')$ can be
expanded in terms of spherical harmonics as
\begin{align}\label{NC3}
\rho(\vec r')=\rho_0(r')Y_{00}(\vec n')+\rho_2(r')Y_{20}(\vec
n')+\rho_4(r')Y_{40}(\vec n')+\cdots
\end{align}
with the multipole components
\begin{align}\label{NC3a}
\rho_l(r)=\int d\vec n\rho(\rb)Y_{l0}(\vec n)\,,
\end{align}
where $\vec{n} =\vec{r}/r$.
Making use of the usual spherical harmonic expansion of
$|\rb_e-\rb|^{-1}$, expression (\ref{v4}) can be written as
\begin{align}\label{NC1}
\la \chi_0^{\lambda}| V | \chi_0^{\lambda}\ra=&-\frac{e^2 Z}{4\pi}\sum_{l=0}^{\infty} \sum_{m=-l}^l \int d\rb'
\bigg( \frac{r_e^l}{r^{l+1}}\Theta(r-r_e) +
\frac{r^l}{r_e^{l+1}}\Theta(r_e-r) \bigg)\notag\\
&\times \rho(\rb') \frac {4\pi}
{2l+1}
 Y_{lm}(\vec{n})Y_{lm}^*(\vec{n}_e).
\end{align}
To integrate over the nuclear angular variables, we
transform $Y_{lm}(\vec{n})$ in Eq. (\ref{NC1}) to the nuclear
coordinate frame
\begin{align}\label{y01}
 Y_{lm}(\vec{n})=\sum_{m'=-l}^{l} Y_{lm'}(\vec{n}')D_{m'm}^{l}(\Phi, \Theta, \Psi).
\end{align}
We have
\begin{align}\label{NC41}
\la \chi_0^{\lambda}| V | \chi_0^{\lambda}\ra=\sum_{k=0}^{\infty} v_{2k}(\rb_e),
\end{align}
where
\begin{eqnarray}\label{NC4}
v_l(\rb_e)&=&-\frac{e^2 Z}{4\pi}
\int \limits_{0}^{\infty}dr r^2 \rho_l(r)\bigg(
\frac{r_e^l}{r^{l+1}}\Theta(r-r_e) +
\frac{r^l}{r_e^{l+1}}\Theta(r_e-r) \bigg)\nonumber\\
&&\times{\frac{4\pi}{2l+1}} \sum_{m=-l}^l Y_{lm}(\vec
n_e)D_{0m}^{l}(\Phi, \Theta, \Psi)\,.
\end{eqnarray}

In the following, we restrict our
 calculations  of nuclear size effects on
atomic binding energies to even-A nuclei with total spin $I=0$ in the ground state.
An extention to non-zero nuclear angular momenta ($I\ne 0$)
can be performed in a similar way.
In the case $I=0$, the interaction potential (\ref{v1}) must be averaged with
the nuclear state  $|IMK\ra=|000\ra$:
\begin{align}\label{VO1}
v(\vec r_e)\equiv\la 000 | V | 000 \ra =& -\frac{e^2 Z}{4\pi}
\sum_{l=0}^\infty \sum_{m=-l}^l \int\limits_0^\infty dr r^2\rho_l(r)
\bigg( \frac{r_e^l}{r^{l+1}}\Theta(r-r_e) +
\frac{r^l}{r_e^{l+1}}\Theta(r_e-r) \bigg) \notag\\
&\times
 \frac{4\pi}{2l+1}Y_{lm}(\vec n_e) \int\limits_0^{2\pi} d\Phi \int\limits_0^{\pi}
d\Theta \sin \Theta  \int\limits_0^{2\pi} d\Psi \notag\\
&\times\frac{1}{\sqrt{8\pi^2}}D_{00}^{0*}(\Phi,\Theta,\Psi)
D_{0m}^{l}(\Phi,\Theta,\Psi)
\frac{1}{\sqrt{8\pi^2}}D_{00}^0(\Phi,\Theta,\Psi)\notag\\
=&-\frac{e^2 Z}{\sqrt{4\pi}} \int\limits_0^\infty dr r^2\rho_0(r)
\bigg( \frac{1}{r}\Theta(r-r_e) + \frac{1}{r_e}\Theta(r_e-r)
\bigg).
\end{align}
Here $\rho_0$ is defined by Eq. (\ref{NC3a}):
\begin{align}
\rho_0(r)=\int d\vec n
\rho(\rb) Y_{00}(\vec n)=\frac{1}{\sqrt{4\pi}}\int d\vec n
\rho(\rb).
\end{align}
In terms of the usual
spherically-symmetric nuclear charge density
\begin{align}
\rho(r)=\frac{1}{4\pi}\int d\vec n \rho(\rb)
\end{align} we obtain
\begin{align}\label{V02}
v(r_e)=-4\pi \alpha Z \int\limits_0^\infty dr r^2\rho(r)  \bigg(
\frac{1}{r}\Theta(r-r_e) + \frac{1}{r_e}\Theta(r_e-r) \bigg) ,
\end{align}
where $\alpha$ is the fine structure constant.
Thus, if we restrict our consideration to the case $I=0$, the summation over $l$
disappears and the interaction potential becomes
spherically-symmetric.
To calculate the energy shift due to the finite-nuclear-size effect one has to solve the Dirac equation with the potential
$v(r)$ given by Eq. (\ref{V02}).

For deformed nuclei the nuclear charge density  is usually described by a modified Fermi model
\begin{align}\label{pho1}
 \rho(\rb)=\frac{N}{1+\exp{[(r-c)/a}]},
\end{align}
 with $\beta$ parameterization of nuclear deformation
\begin{align}\label{pho2}
 c=c_0
 \big(1+\sum^\infty_{l=1}\sum^{l}_{m=-l}\beta_{lm}Y_{lm}(\Theta,\varphi)\big)
\end{align}
consistent with the normalization condition:
\begin{align}
\int d^3r\rho(\rb)=1.\notag
\end{align}
Assuming axial symmetry and considering only quadrupole and hexadecapole nuclear deformation, the expression
(\ref{pho2}) reduces to
\begin{align}\label{pho10}
 c=c_0\big(1+\beta_{20}Y_{20}+\beta_{40}Y_{40}\big).
\end{align}

Before turning to the numerical evaluations of the nuclear-size correction for some ions of experimental interest,
we also present approximate analytical formulas that explicitely take the nuclear deformation into account.

\section{Approximate analytical formulas for the energy shift}

According to the method of Ref. \cite{Sh93} the calculation of the one-electron
finite-nucleus-size correction $\dE$ for an arbitrary
nuclear model can be reduced to the calculation of $\dE$ for the model of a
homogeneously charged sphere with an effective radius $R$. To a high degree of accuracy,
the effective nuclear radius for $j=1/2$ states is given by \cite{Sh93}
\begin{align}\label{R1}
    R=\Bigg\{ \frac{5}{3} \la & r^{2} \ra
       \Bigg[1-\frac{3}{4}(\az)^2 \Bigg(\frac{3}{25} \frac{\la r^{4}\ra}
  {\la r^{2} \ra^{2}}
    -\frac{1}{7}\Bigg)\Bigg]\Bigg\}^{1/2},
\end{align}
where
\begin{align}
\la r^n \ra=\frac{1}{4\pi}\int d\rb\rho(\rb)r^n
=\int\limits_0^\infty dr r^{n+2}\rho(r) .
\end{align}
Then $\dE$ can be evaluated using the following
approximative formulas \cite{Sh93}:
\begin{align}\label{E1}
 \dE_{ns_{\frac{1}{2}}}=&\frac{(\az)^2}{10n}(1+(\az)^2 f_{ns_{\frac{1}{2}}}
  (\az))\bigg(2\frac{\az R}{n\lambda_{C}}\bigg)^{2\gamma}mc^2,\\
\label{E2}
 \dE_{np_{\frac{1}{2}}}=&\frac{(\az)^4}{40}\frac{n^2-1}{n^3}(1+(\az)^2f_{np_{\frac{1}{2}}}
 (\az))\bigg(2\frac{\az R}{n\lambda_{C}}\bigg)^{2\gamma}mc^2,\\
\label{E3}
 f(\az)=&b_0+b_1(\az)+b_2(\az)^2+b_3(\az)^3.
\end{align}
Here $n$ is the principal quantum number,
$\lambda_{C}=\hbar/mc$, and $\gamma=\sqrt{1-(\az)^2}$. The
coefficients $b_0-b_3$ for a number of states are given in Ref. \cite{Sh93}.
Formulas (\ref{R1})-(\ref{E3}) allows one to calculate $\dE$
in the range $Z=1-100$ with a relative accuracy of $0.2\%$. For more
precise formulas we refer to Refs. \cite{Sh93,Sh02}.

For the deformed Fermi distribution given by Eqs. (\ref{pho1}), (\ref{pho10}) within the $\be^4$ and $\bee^2$  approximation (as a rule, $\be^4\sim\bee^2$), we derive
\begin{align}\label{pho3}
 N=&\frac{3}{4\pi c_0^3}\bigg\{1+\bigg(\frac{\pi a}{c_0}\bigg)^2+
\frac{3}{4\pi}\bigg(1+\frac{3}{7\sqrt{\pi}}\bee\bigg)\be^2
+\frac{1}{28\pi}\sqrt{\frac{5}{\pi}}\be^3
+\frac{3}{4\pi}\bee^2\bigg\}^{-1} , \\
\label{pho4} \la r^2\ra=&\frac{4}{5}\pi
Nc_0^5\bigg\{1+\frac{10}{3}\bigg(\frac{\pi a}{c_0}\bigg)^2
+\frac{7}{3}\bigg(\frac{\pi a}{c_0}\bigg)^4   \notag\\
&+\frac{5}{2\pi}\bigg[1+\frac{9}{7\sqrt{\pi}}\bee+\bigg(\frac{\pi
a}{c_0}\bigg)^2\bigg(1+\frac{3}{7\sqrt{\pi}}\bee\bigg)\bigg]\be^2\notag\\
&+\frac{5}{42\pi}\sqrt{\frac{5}{\pi}}\bigg[3+\bigg(\frac{\pi
a}{c_0}\bigg)^2\bigg]\be^3 +\frac{75}{112\pi^2}\be^4
+\frac{5}{2\pi}\bigg[1+\bigg(\frac{\pi
a}{c_0}\bigg)^2\bigg]\bee^2\bigg\},\\
\label{pho5} \la r^4\ra=&\frac{4}{7}\pi
Nc_0^7\bigg\{1+7\bigg(\frac{\pi a}{c_0}\bigg)^2+
\frac{49}{3}\bigg(\frac{\pi
a}{c_0}\bigg)^4+\frac{31}{3} \bigg(\frac{\pi a}{c_0}\bigg)^6 \notag\\
&+\frac{21}{4\pi}
\bigg[1+\frac{15}{7\sqrt{\pi}}\bee
+\bigg(\frac{10}{3}+\frac{30}{7\sqrt{\pi}}\bee\bigg)\bigg(\frac{\pi a}{c_0}\bigg)^2
+\bigg(\frac{7}{3}+\frac{1}{\sqrt{\pi}}\bee\bigg)\bigg(\frac{\pi a}{c_0}\bigg)^4\bigg]\be^2 \notag\\
&+ \frac{5}{4\pi}\sqrt{\frac{5}{\pi}}\bigg[1+2\bigg(\frac{\pi
a}{c_0}\bigg)^2
+\frac{7}{15}\bigg(\frac{\pi a}{c_0}\bigg)^4 \bigg]\be^3\notag\\
&+\frac{75}{16\pi^2}\bigg[1+\bigg(\frac{\pi a}{c_0}\bigg)^2\bigg]
\be^4
+\frac{21}{4\pi}\bigg[1+\frac{10}{3}\bigg(\frac{\pi
a}{c_0}\bigg)^2+\frac{7}{3}\bigg(\frac{\pi a}{c_0}\bigg)^4\bigg]
\bee^2\bigg\}\,.
\end{align}
Expanding $\la r^2\ra$ and $\la r^4\ra$ in terms of the $\beta$
parameters and keeping the two lowest-order terms yields
\begin{align}\label{pho6}
\la r^2\ra=&\frac{1}{5}(3c_0^2+7\pi^2a^2)+\frac{7c_0^2 +3(\pi
a)^2}{1+(\frac{\pi a}{c_0})^2}\frac{3}{20\pi}\be^2 +
\frac{9c_0^2+(\pi a)^2}{1+(\frac{\pi
a}{c_0})^2}\frac{3}{140\pi}\sqrt{\frac{5}{\pi}}\be^3,\\
\label{pho7} \la r^4\ra=&\frac{1}{7}(3c_0^4 +
18\pi^2a^2c_0^2+31\pi^4a^4)+\frac{9c_0^4+26\pi^2a^2c_0^2+9\pi^4a^4}{1+(\frac{\pi
a}{c_0})^2}\frac{3}{14\pi}\be^2 \notag\\
&+\frac{17c_0^4+32\pi^2a^2c_0^2+3\pi^4a^4}{1+(\frac{\pi a}{c_0})^2}
\frac{3}{98\pi}\sqrt{\frac{5}{\pi}}\be^3.
\end{align}
In the limit, where $\be$ tends to zero, the ordinary Fermi
distribution is recovered. Substituting Eqs.
(\ref{pho6})-(\ref{pho7}) into formulas
(\ref{R1})-(\ref{E2}), one immediately finds $\dE$ for a hydrogenlike atom with a deformed
nucleus, provided the parameters $c_0$, $a$, and $\be$ are known.

To study the role of nuclear deformation in calculations of the
finite-nuclear-size correction, let us consider the energy difference for two isotopes.
Since this difference can be approximated as \cite{Sh93}
\begin{align}\label{E04}
\delta E=\dE_2-\dE_1\simeq2\gamma(\delta R/R)\dE,
\end{align}
we have to find the dependence of $\delta R/R$ on
variations of the nuclear charge distribution parameters. Assuming that the
value $1/N$, which determines the
nuclear volume at $a=0$, is proportional to atomic number $A$, we derive
 \begin{align}\label{E04aa}
\frac{\delta R}{R}\simeq
\frac{1}{3}\frac{\delta A}{A}
+\frac{1}{2}\frac{\pi^2\delta(a^2)}{\langle r^2\rangle}
+\frac{5}{8\pi}\delta(\be^2),
 \end{align}
where the first term is due to an increase of the nuclear volume,
the second one results from a change of the parameter $a$, and
the third one represents nuclear deformation \cite{wil53,wu69}.
If the spherically-symmetric nucleus is considered as a reference
($\delta(\be^2) = \be^2 $), and the parameter $a$ is the same
for both isotopes, we get
\begin{align}\label{E04a}
\frac{\delta R}{R}\simeq \frac{1}{3}\frac{\delta A}{A} +
\frac{5}{8\pi}\be^2.
\end{align}
This formula gives a simple
way to determine the nuclear deformation parameter $\be$, provided the isotope
 shift is known, e.g., from experiment.

Alternatively, considering $\langle r^2\rangle ^{1/2}$, $a$, and $\be$ as
free independent parameters, we obtain
 \begin{align}\label{E04b}
\frac{\delta R}{R}\simeq
\frac{\delta \langle r^2\rangle ^{1/2}}{\langle r^2\rangle ^{1/2}}
 -\frac{3}{70}(\az)^2\frac{\pi^2\delta(a^2)}{\la r^2\ra}
-\frac{3}{56\pi}(\az)^2\delta(\be^2).
 \end{align}
This formula shows that, to a good accuracy, the isotope shift is
determined by the change of the rms radius.

\section{Nuclear size correction to the binding energies in $^{238}\textrm{U}^{91+}$ and $^{238}\textrm{U}^{89+}$}

In this section  the formulation given above is applied to deduce a new value for the rms radius in $^{238}\textrm{U}$ and, with this value, to revise theoretical predictions for the ground state Lamb shift in $^{238}\textrm{U}^{91+}$ and for the $2p_{1/2}-2s$ transition energy in $^{238}\textrm{U}^{89+}$.

Compilation of the rms values \cite{An99,An04,Fr95} employed experimental data for nuclear charge distribution parameters obtained by various experimental methods. In case of $^{235,238}\textrm{U}$ the most recent compilation by Angeli \cite{An99,An04} includes data from elastic electron scattering \cite{Cr77}, muonic atom X-rays \cite{Cl78,Zu84}, X-ray isotope shifts \cite{Br65,El96,El98}, and optical isotope shifts \cite{An92}.
Since in Ref. \cite{Cl78}  the experimental data are given in terms of the parameters $a$, $c_0$, $\be$, and $\bee$,
one should first evaluate the corresponding rms values.
In Refs. \cite{An99,An04} this was achieved based on formulas which only partly account
for the deformation effect. In the present work we improved the Angeli's evaluation
employing  formulas
(\ref{pho3})-(\ref{pho4}) as well as the direct numerical calculations.
As a result, we obtained
the  $\la r^2\ra^{1/2}$ values
which are close enough to those from the other sources \cite{Cr77,Zu84}.
In case of $^{238}$U,
the compillation of the improved data for $\la r^2\ra^{1/2}$ and
 the  $\delta \la r^2\ra$ data from Refs. \cite{Br65,El96,El98,An92},
 performed by Angeli \cite{An07}, yields
$\la r^2\ra^{1/2}=5.8569(33)$ fm.
This value differs from the corresponding value from the previous compillation,
$\la r^2\ra^{1/2}=5.8507(72)$ fm \cite{An04}.
As to the other nuclear-charge-distribution parameters,
in accordance with the available experimental data \cite{Cl78,Zu84}, we use
$a=0.50(5)$ fm, $\be=0.27(1)$, and $\bee=0.05(10)$ assuming rather
conservative errors bars.
 These parameters
differ from those employed in similar calculations by Blundell \textit{et al.}
\cite{Bl90} and by Ynnerman \textit{et al.} \cite{Yn94}, who adopted exclusively the data of the muonic X-ray experiment \cite{Zu84}.

The finite-nuclear-size correction is obtained by solving the Dirac equation with the potential (\ref{V02}) and taking the difference between the energies for the extended and the point-charge nucleus.
In order to investigate the importance of the nuclear
deformation effect, the calculations of the finite-nuclear-size correction are also
performed using a spherically-symmetric nuclear charge distribution with the same rms value or
with the same nuclear volume.
The results of these calculations are compared with each other in Table \ref{Table_U1}.
In addition to the direct numerical (N) calculations, the analytical (A) results obtained
by formulas (\ref{R1})-(\ref{pho5}), which provide a 0.2\% accuracy, are presented as well.
 As one can see from the table, if the rms value is kept to be the same, the nuclear deformation provides a $0.06\%$ energy shift.
If the nuclear volume is constant, the energy shift amounts to about $2\%$.
It can also be seen that the energy shifts
obtained by analytical formulas (\ref{E04})-(\ref{E04b}) are in a reasonable agreement with the exact numerical results. We note also that the effect of  hexadecapole deformation $(\sim \beta_{40})$
 is extremely small  for $^{238}\textrm{U}$, provided the rms radius is kept to be constant.

\begin{turnpage}
\begin{table}
 \caption{The exact numerical (N) and approximate analytical (A) results for
the finite-nuclear-size correction to the energies of $1s$, $2s$, and $2p_{1/2}$ states of  $^{238}\textrm{U}^{91+}$ ($\la r^2\ra^{1/2}=5.8569(33)$ fm, $a=0.50(5)$ fm, $\be=0.27(1)$, and $\bee=0.05(10)$), in eV. The results for a deformed (D) nucleus are compared with the results obtained for a spherically-symmetric nuclear model with $(1)$ the same value of the rms value $(\la r^2\ra^{1/2}=\la r^2\ra^{1/2}_\textrm{D})$ or with $(2)$ the same nuclear volume $(1/N=(1/N)_\textrm{D})$.}
\label{Table_U1}
 \linespread{1}
   \begin{center}
    \begin{tabular}{cccccccccccccc}
     \hline
     \hline
Nuclear &$\la r^2\ra^{1/2}$&$\la r^4\ra^{1/4}$&$a$&$\be$&$\bee$&$c_0$&
$1s$&$2s$&$2p_{1/2}$&$2p_{1/2}$&
Method\\
model&&&&&
&&&&&-$2s$\\
&(fm)&(fm)&(fm)&&&(fm)&(eV)&(eV)&(eV)&(eV)&\\
     \hline

Def. nuc. &  5.8569(33)&  6.2384 &   0.50(5)&   0.27(1)&   0.05(10)&   7.0140 &
           198.54(19)& 37.714(34)&  4.410(4)&   -33.304(30) &N\\
          & & & & & & &
           198.39& 37.651&  4.412 &-33.239  &A\\

$(1)$ Sph. sym. &  5.8569&  6.2088&  0.50&   0.00&   0.00&   7.1704& 
            198.68  &37.740&  4.413&   -33.327 &N\\
$\la r^2\ra^{1/2}=\la r^2\ra^{1/2}_\textrm{D}$
          & & & & & & &
            198.61  &37.692&  4.417& -33.275&  A\\

$(2)$ Sph. sym. &  5.7805&  6.1303&   0.50&     0.00&   0.00&  7.0663&

          194.90&  37.025&  4.328 &-32.696 &N\\
$1/N=(1/N)_\textrm{D}$
          & & & & & & &
          194.77&  36.963&  4.331 &-32.632  &A\\

     \hline
     \hline
     \end{tabular}
   \end{center}
\end{table}
\end{turnpage}

Thus to calculate the nuclear size correction for $^{238}\textrm{U}^{91+}$ to a $0.1\%$ accuracy one needs to account for the nuclear deformation effect.
Finally, the nuclear-size corrections for $^{238}\textrm{U}^{91+}$ are $\Delta E (1s)=198.54(19)$ eV, $\Delta E (2p_{1/2}-2s)=-33.304(30)$ eV,
and $\Delta E (2p_{3/2}-2s)=-37.714(34)$ eV.

In the last compilations of the ground-state Lamb shift in $^{238}\textrm{U}^{91+}$ \cite{Yer03,Sha06} and the $2p_{1/2}-2s$ transition energy in $^{238}\textrm{U}^{89+}$ \cite{Ye06,Sh06a} the total theoretical uncertainties were mainly determined by the finite-nuclear-size corrections. The new values for these corrections obtained in the present work provide significant improvements of the theoretical predictions for both H- and Li-like uranium. In Table \ref{U_1s} we present individual contributions to the $1s$ Lamb shift in $^{238}\textrm{U}^{91+}$. The uncertainty of the total theoretical value, $463.99(39)$ eV, is now mainly determined by uncalculated two-loop QED corrections, in particular,
the mixed vacuum-polarization self-energy contribution \cite{zsch_02}.
The obtained result is in a good agreement with the recent experiment \cite{Gum05}.
\begin{table}
 \caption{Individual contributions to the ground-state Lamb shift in $^{238}\textrm{U}^{91+}$, in eV.}
\label{U_1s}
   \begin{center}
    \begin{tabular}{lr@{}lr}
     \hline
     \hline
Contribution&                   Value&     &  Reference\\
Finite nuclear size&            198.&54(19) & This work\\
First-order QED&            266.&45&         \cite{Mo98} \\
Second-order QED&            -1.&26(33)&         \cite{Yer03}\\
Nuclear recoil      &            0.&46&         \cite{Sh98}\\
Nuclear polarization&    -0.&20(10)&         \cite{Pl95,Ne96}\\
Total theory &           463.&99(39)&         \\
Experiment   &           460.&2(4.6)&         \cite{Gum05}\\
     \hline
     \hline
     \end{tabular}
   \end{center}
\end{table}

Table \ref{U_Lamb} presents individual contributions to the $2p_{1/2}-2s$ transition energy in $^{238}\textrm{U}^{89+}$. Compared to Refs. \cite{Ye06,Sh06a}, it contains the new value for the nuclear-size correction and the new value for the three- and more photon effects. The latter correction was evaluated within the Breit approximation employing the large-scale configuration-interaction Dirac-Fock-Sturm (CI-DFS) method \cite{Tup03,Tup05}. The procedure successfully used  for Li-like scandium \cite{Ko07} was applied here as well. For uranium, we report a good agreement with the previous evaluations of this correction \cite{Zh00,An01,Yer07}. The uncertainty ascribed to this correction incorporates all three- and more photon effects which are beyond the Breit approximation.  The entry labeled "Screened QED" represents the sum of the lowest-order self-energy and vacuum-polarization screening diagrams \cite{Ar99,Yer99}.

\begin{table}
 \caption{Individual contributions to the $2p_{1/2}-2s$ transition energy in $^{238}\textrm{U}^{89+}$, in eV.}
\label{U_Lamb}
   \begin{center}
    \begin{tabular}{lr@{}lr}
     \hline
     \hline
Contribution &            Value& & Reference\\
One-electron nuclear size&  -33.&30(3) & This work\\
One-photon exchange&        368.&83   & This work   \\
One-electron first-order QED&   -42.&93& \cite{Mo98} \\
Two-photon exchange within the Breit approx.&            -13.&54& \cite{Ye00}\\
Two-photon exchange beyond the Breit approx.&              0.&17& \cite{Ye00}\\

Screened QED&            1.&16& \cite{Ar99,Yer99}\\
One-electron second-order QED&            0.&22(6)& \cite{Ye06}\\
Three- and more photon effects&   0.&14(7)& This work\\
Nuclear recoil      &            -0.&07& \cite{Ar95}\\
Nuclear polarization&    0.&03(1)& \cite{Pl95,Ne96}\\
Total theory &           280.&71(10)& \\
Experiment   &           280.&645(15)& \cite{Be05}\\
Experiment   &           280.&59(10)& \cite{Sc91}\\
Experiment   &           280.&52(10)& \cite{Br03}\\
     \hline
     \hline
     \end{tabular}
   \end{center}

\end{table}

Table \ref{U_Lamb} shows that now, after our revision of the finite-nuclear-size correction, the total theoretical uncertainty is mainly influenced by higher-order QED effects.
The total theoretical value of the transition energy, $280.71(10)$ eV, agrees well with the most precise experimental value,
280.645(15) eV \cite{Be05}. Comparing the first- and second-order QED contributions with the total theoretical uncertainty, we conclude that the present status of the theory and experiment for Li-like uranium provides a test of QED on a $0.2\%$ level to first order in $\alpha$ and on a $6.5\%$ level to second order in $\alpha$.

\section{Isotope shift of the $2p_j-2s$ transition energies for $^{142}\textrm{Nd}^{57+}$ and $^{150}\textrm{Nd}^{57+}$}

In this section  we evaluate the isotope shift of the $2p_{j}-2s$ transition energies for the isotopes A$=142$ and A$=150$ of Li-like $^\textrm{A}\textrm{Nd}^{57+}$, where the $^\textrm{150}\textrm{Nd}$ nucleus is strongly deformed $(\be=0.28(5)$, see, e.g., Ref. \cite{Fr95}).
To date, there are about 20 publications, where the mean-square charge
radius difference $\delta\la r^2\ra$ for these isotopes is reported (see Refs. \cite{Brpu,Fr95,An99,An04} and references therein).
Apart from some outliers, the majority of the
 experimental data cover a range from about $^{150,142}\delta\la r^2\ra=1.18$ fm$^2$ to $^{150,142}\delta\la r^2\ra=1.38$ fm$^2$. For this reason, we perform calculations of the isotope shift for the entire range of $^{150,142}\delta\la r^2\ra$, from $1.18$ to $1.38$ fm$^2$. With these data, one can easily find the value of $^{150,142}\delta\la r^2\ra$ from the experimental value of the isotope shift \cite{Brpu}.

The isotope shift is given by the sum of the field shift, which is due to the finite-nuclear-size effect, and the mass shift, which is determined by the nuclear recoil effect. To evaluate the field shift we used the large-scale CI-DFS method \cite{Tup03,Tup05}, with the Breit interaction included. The spherically-symmetric $^{142}\textrm{Nd}$ nucleus served as a reference with the rms radius of $\la r^2\ra^{1/2}=4.9118$ fm from the  compillation by Angeli \cite{An04}. The other nuclear parameters are taken to be $a=0.52(2)$ fm for both isotopes, $\be=0$ for $^{142}\textrm{Nd}$ and $\be=0.28(5)$ for $^{150}\textrm{Nd}$ \cite{Fr95}. We note that variations of these parameters within their error bars do not affect the isotope shifts at the accuracy level considered.

The full relativistic theory of the nuclear recoil effect can be formulated only in the framework of QED \cite{Sh98a}. To evaluate the recoil effect within the lowest-order relativistic approximation one can use the operator \cite{Sh85,Pa87}:
\begin{equation}\label{recoil}
  H_M = \frac{1}{2M} \sum_{i,j} \left[ \boldsymbol{p}_i\cdot\boldsymbol{p}_j
    - \frac{\az}{r_i} \left( \albi+\frac{(\albi\cdot\boldsymbol{r}_i)\boldsymbol{r}_i}
    {r^{2}_{i}} \right) \cdot\boldsymbol{p}_j \right],
\end{equation}
where $M$ is the nuclear mass and $\boldsymbol{p}_i$ is the momentum operator
acting on the i-th electron. The expectation value of $H_M$ on the many-electron
wave function of the system, obtained by the CI-DFS method, yields the recoil correction
to the energy levels to all orders in $1/Z$ within the $(\az)^4{m^2}/{M}$ approximation.
The recoil correction which is beyond this approximation is termed as the QED recoil effect.
For the $2p_{1/2}-2s$ $(2p_{3/2}-2s)$ transition the mass shift comprises of 2.44 meV (2.53 meV) from averaging the nonrelativistic part of the recoil operator (the first term in Eq. (\ref{recoil})) with the relativistic many-electron wave function, $-1.14$ meV ($-1.03$ meV) from the relativistic part (the second term in Eq. (\ref{recoil})), and of $0.33$ meV (0.30 meV) from the QED recoil effect \cite{Ar95, JPB28_5201}.
The recoil correction of the next order in $m/M$ is negligible in the case under consideration. Finally, the total mass shift sums up to 1.63 meV for the $2p_{1/2}-2s$ transition and to 1.80 meV for the $2p_{3/2}-2s$ transition.

Next, one should account for the influence of the nuclear size variation on the one-loop QED corrections. Using comprehensive tabulations for the nuclear-size correction to the self-energy contribution \cite{Be98} and evaluating the corresponding effect on the vacuum-polarization contribution, we derive $0.2$ meV for the QED correction to the isotope shifts under consideration.

Finally, we have to consider the nuclear polarization (NP) effect. This correction is determined by the electron-nucleus interaction diagrams in which the intermediate states of the nucleus are excited. This effect was evaluated for a number of ions in Refs. \cite{Pl95,Ne96}. Since the NP correction is most sizeable for deformed nuclei, we estimated it for $^\textrm{150}\textrm{Nd}$ taking into account the transition to the first excited $2^+$ state at $130.21$ keV only. Taking the nuclear transition probability from Ref. \cite{Ra01} and evaluating the sum over intermediate electron states numerically as well as analytically according to formulas derived in Ref. \cite{Ne96}, we obtain $0.3(3)$ meV for the nuclear polarization contribution to the isotope shift for both transitions.

The results of our calculations are presented in Tables \ref{Table_nd_is1} and  \ref{Table_nd_is2} for the $2p_{1/2}-2s$ and $2p_{3/2}-2s$ transitions, respectively. With the numbers compiled in
these tables, one can easily deduce the nuclear mean-square charge difference $\delta \la r^2\ra$, provided the isotope shift is known from experiment \cite{Brpu}. In addition, using formula (\ref{E04a}) one can derive the quadrupole deformation parameter  $\be$ to an accuracy of about $20-30\%$.

\begin{table}
 \caption{Isotope shift for the $2p_{1/2}-2s$ transition in Li-like $^{150,142}\textrm{Nd}^{57+}$, in eV. The field shift includes one-electron Dirac, electron-correlation, and Breit-interaction contributions. The mass shift incorporates nonrelativistic, relativistic, and QED recoil effects. The QED correction represents the sum of one-loop self-energy and vacuum-polarization contributions.}
\label{Table_nd_is1}
 \linespread{1}
   \begin{center}
    \begin{tabular}{cccccc}
     \hline
     \hline
$\delta \la r^2 \ra$ & Field shift & Mass shift & QED & Nuc. pol. & Total\\
(fm$^2)$&&&\\
     \hline
   1.180& -0.0366 & 0.0016 & 0.0002 & 0.0003 &   -0.0345\\
   1.200& -0.0372 & 0.0016 & 0.0002 & 0.0003 &   -0.0351\\
   1.220& -0.0379 & 0.0016 & 0.0002 & 0.0003 &   -0.0358\\
   1.240& -0.0385 & 0.0016 & 0.0002 & 0.0003 &   -0.0364\\
   1.260& -0.0391 & 0.0016 & 0.0002 & 0.0003 &   -0.0370\\
   1.280& -0.0397 & 0.0016 & 0.0002 & 0.0003 &   -0.0376\\
   1.300& -0.0403 & 0.0016 & 0.0002 & 0.0003 &   -0.0382\\
   1.320& -0.0410 & 0.0016 & 0.0002 & 0.0003 &   -0.0389\\
   1.340& -0.0416 & 0.0016 & 0.0002 & 0.0003 &   -0.0395\\
   1.360& -0.0422 & 0.0016 & 0.0002 & 0.0003 &   -0.0401\\
   1.380& -0.0428 & 0.0016 & 0.0002 & 0.0003 &   -0.0407\\
     \hline
     \hline
     \end{tabular}
   \end{center}
\end{table}

\begin{table}
 \caption{Isotope shift for the $2p_{3/2}-2s$ transition in Li-like $^{150,142}\textrm{Nd}^{57+}$, in eV. The field shift includes one-electron Dirac, electron-correlation, and Breit-interaction contributions. The mass shift incorporates nonrelativistic, relativistic, and QED recoil effects. The QED correction presents the sum of one-loop self-energy and vacuum-polarization contributions.}
\label{Table_nd_is2}
 \linespread{1}
   \begin{center}
    \begin{tabular}{cccccc}
     \hline
     \hline
$\delta \la r^2 \ra$ & Field shift & Mass shift & QED & Nuc. pol. & Total\\
(fm$^2)$&&&\\
     \hline
   1.180& -0.0379 & 0.0018 & 0.0002 & 0.0003 &   -0.0353\\
   1.200& -0.0385 & 0.0018 & 0.0002 & 0.0003 &   -0.0362\\
   1.220& -0.0392 & 0.0018 & 0.0002 & 0.0003 &   -0.0369\\
   1.240& -0.0398 & 0.0018 & 0.0002 & 0.0003 &   -0.0375\\
   1.260& -0.0404 & 0.0018 & 0.0002 & 0.0003 &   -0.0381\\
   1.280& -0.0411 & 0.0018 & 0.0002 & 0.0003 &   -0.0388\\
   1.300& -0.0417 & 0.0018 & 0.0002 & 0.0003 &   -0.0394\\
   1.320& -0.0424 & 0.0018 & 0.0002 & 0.0003 &   -0.0401\\
   1.340& -0.0431 & 0.0018 & 0.0002 & 0.0003 &   -0.0408\\
   1.360& -0.0437 & 0.0018 & 0.0002 & 0.0003 &   -0.0414\\
   1.380& -0.0443 & 0.0018 & 0.0002 & 0.0003 &   -0.0420\\
     \hline
     \hline
     \end{tabular}
   \end{center}
\end{table}

\section{Conclusion}

The finite-nuclear-size correction to the binding energies in heavy ions has been studied in this paper. In the general case of a deformed nucleus, approximate analytical formulas for this effect have been derived and direct numerical calculations have been performed. In the special case of $^{238}\textrm{U}$ the study has been employed to revise the nuclear-charge-distribution parameters and to recalculate the binding energies in H- and Li-like uranium. As the result, the largest theoretical uncertainties for
the ground-state Lamb shift in $^{238}\textrm{U}^{91+}$ and for the $2p_{1/2}-2s$ transition energy in $^{238}\textrm{U}^{89+}$  have been removed. Now the total theoretical accuracy is mainly restricted by higher-order QED effects. Tables \ref{U_1s} and \ref{U_Lamb} demonstrate that our theoretical
results agree well within the error bars with the most precise experimental data.

We have also evaluated the isotope shift of the $2p_j-2s$ transition energies for $^{142}\textrm{Nd}^{57+}$ and $^{150}\textrm{Nd}^{57+}$ for different values of the mean-square nuclear charge difference $\delta\la r^2 \ra$. The calculation of the field shift takes into account electron-correlation, Breit-interaction, and QED effects. The mass shift is evaluated within a full QED treatment. The nuclear-polarization  correction is also estimated. The data obtained allow one to extract the $\delta\la r^2 \ra$ value from the corresponding experiment.

\acknowledgments
Valuable communications with I. Angeli are gratefully acknowledged.
This work was supported in part by INTAS
(Grant No. 06-1000012-8881), RFBR (Grant No. 07-02-00126), DFG, DAAD, and GSI.
Y.S.K. and O.V.A. acknowledge the support by the "Dynasty" foundation.


\end{document}